# Multiobjective Direct Policy Search Using Physically Based Operating Rules in Multireservoir Systems

Josias Ritter[1]; Gerald Corzo[2]; Dimitri P. Solomatine[3]; and H. Angarita[4]

[1]Center of Applied Research in Hydrometeorology, Universitat Politècnica de Catalunya, Jordi Girona 1-3 (C4), Barcelona 08034, Spain (corresponding author). ORCID: https://orcid.org/0000-0002 -3833-5450. Email: ritterjosias@gmail.com
[2]Senior Lecturer, Chair Group of Hydroinformatics, IHE Delft Institute for Water Education, Westvest 7, Delft 2611AX, Netherlands.
[3]Professor and Head, Chair Group of Hydroinformatics, IHE Delft Institute for Water Education, Westvest 7, Delft 2611AX, Netherlands; Professor (Zero-Nominated), Water Resources Section, Delft Univ. of Technology, Mekelweg 5, Delft 2628CD, Netherlands; Guest Researcher, Flood Hydrology Lab, Water Problems Institute of RAS, Gubkina St. 3, Moscow 119333, Russia.
[4]Research Fellow, Latin America Centre, Stockholm Environment Institute, Calle 71 No. 11-10 Of. 801, Bogotá 110231, Colombia. ORCID: https://orcid.org/0000-0001-7089-2014

## Abstract

This study explores the ways to introduce physical interpretability into the process of optimizing operating rules for multireservoir systems with multiple objectives. Prior studies applied the concept of direct policy search (DPS), in which the release policy is expressed as a set of parameterized functions (e.g., neural networks) that are optimized by simulating the performance of different parameter value combi- nations over a testing period. The problem with this approach is that the operators generally avoid adopting such artificial black-box functions for the direct real-time control of their systems, preferring simpler tools with a clear connection to the system's physics. This study addresses this mismatch by replacing the black-box functions in DPS with physically based parameterized operating rules, for example by directly using target levels in dams as decision variables. This leads to results that are physically interpretable and may be more acceptable to operators. The methodology proposed in this work is applied to a network of five reservoirs and four power plants in the Nechí catchment in Colombia, with four interests involved: average energy generation, firm energy generation, flood hazard, and flow regime alteration. The release policy is expressed depending on only 12 parameters, which significantly reduces the computational complexity compared to existing approaches of multiobjective DPS. The resulting four-dimensional Pareto-approximate set offers a variety of operational strategies from which operators may choose one that corresponds best to their preferences. For demonstration purposes, one particular optimized policy is selected and its parameter values are analyzed to illustrate how the physically based operating rules can be directly interpreted by the operators.

*Author keywords:* Multiobjective reservoir optimization; Multireservoir systems; Direct policy search; Parameterization simulation optimization; Policy myopia.

## Introduction

One of the main challenges in river basin management has always been finding a balance between competing water demands. Reservoirs play a key role in this task. Traditionally, reservoirs were designed and operated to satisfy mainly one of several interests (i.e., hydropower or flood control; Lund and Guzman 1999), but for several decades understanding has become widely spread that it is a multifactor multiobjective problem (Yeh 1985). Additionally, driving factors such as climate change, population growth, and increasing environmental standards pose additional pressures on river basins, resulting in stronger exploitation and an increasing number of interests involved (McDonald et al. 2011). As opposed to expensive and often undesirable infrastructural interventions (e.g., the construction of additional reservoirs), the increasing pressures can be addressed by using existing infrastructure more efficiently (Gleick and Palaniappan 2010; Whateley et al. 2014). Revising operating rules is an immediate measure that requires comparatively little resources.

There are numerous techniques to optimize reservoir operations; see Labadie (2004) for the most common approaches. Around 30–40 years ago, multiobjective problems were approached by means of multicriteria decision analysis (MCDA); for a review of the most common MCDA methods, see Chankong and Haimes (1983). Multicriteria decision analysis methods assume a priority ranking between objectives based on "a priori" (i.e., prior to optimization) preferences of a decision maker (DM), defining weights that are applied to the individual objectives and then aggregating them; see Delipetrev et al. (2017) and Bai et al. (2017) for recent applications. The multiobjective problem is thus converted into a single-objective problem that (through the application of any single-objective optimization technique) allows for identification of one optimal compromise solution. Multicriteria decision analysis methods are based on strong assumptions, among which is the necessity of a well-defined understanding of the DM's preferences (Haimes and Hall 1977). Defining the preferences of the DM "a priori" may lead to the failure to explore the entire set of tradeoffs between interests, and therefore decreases the possibility to discover decision-relevant alternatives. This problem, also known as policy myopia (Giuliani et al. 2014b), biases the decisions and the resulting operating policies (Hogarth 1981).

In order to avoid these biases, a great deal of research has been carried out to develop "a posteriori" techniques that explore the full Pareto-approximate set of solutions, taking into account multiple objectives, before including the DM's preferences and making final choices. One approach is to employ the "a priori" methods described above, varying the weights that express the preferences of the DM, which results in a variety of solutions that jointly form the Pareto-approximate set (Soncini-Sessa et al. 2007). Another widely used approach is to employ multiobjective evolutionary algorithms (MOEAs); for a review of the most common MOEAs, see Reed et al. (2013) and Maier et al. (2014). These algorithms belong to the class of randomized search algorithms, and the main idea is to follow a cycle of random sampling of decision variable vectors (representing the parameters of a policy), combining the best solutions and progressively eliminating the worst ones. The result is a Pareto-approximate set of decision variable vectors, none of which is dominated (i.e., there are no solutions that are better than any other with respect to all objective functions).

Formerly the optimization of release policies toward multiple objectives was computationally very difficult, especially the idea of optimizing specific decisions, like individual releases at each time step of the control horizon, which usually involves a high number of decision variables; for a review of approaches following this idea, see Lin and Rutten (2016). An alternative approach that comes with a much lower computational cost is the optimization of general reservoir management rules that are valid over the entire control horizon and generate releases depending on external drivers (e.g., inflows). Rules of this kind are often prescribed by the authorities. This latter approach is capable of finding solutions that are—in terms of the quality of the resulting decisions—comparable to the former approach, and the number of decision variables can be made comparably low (Koutsoyiannis and Economou 2003). This less demanding computational approach was adopted by a variety of methods (Labadie 2004); a widely used method is the direct policy search (DPS) (Williams 1992; Rosenstein and Barto 2001), which is also referred to as parameterization-simulation-optimization in water resources literature (Koutsoyiannis and Economou 2003). The direct policy search expresses a release policy using a given family of parameterized functions (Deisenroth et al. 2013). It was also extended to multiobjective problems (Biglarbeigi et al. 2014) and MOEAs were used to solve it. For instance, Giuliani et al. (2016) found radial basis functions (RBF) to be particularly useful for this task. They employed a MOEA for optimization of parameters in the RBF with respect to two objectives and, understandably, named it Evolutionary Multiobjective Direct Policy Search (EMODPS).

The evolutionary multiobjective direct policy search approximates a Pareto set consisting of individual parameterizations of operational policies. Each policy can control the system; choosing "the best one" depends on the DM's preferences regarding the objectives. However, these policies are represented by mathematical functions with parameters that do not have physical meaning and, in reality, reservoir operators are often reluctant to use such black-box functions for direct real-time control of their systems (Teegavarapu and Simonovic 2001; Celeste and Billib 2009; Giuliani et al. 2014b). They prefer making their decisions based on tools that have an apparent connection to the operating conditions and physics of the systems they operate (e.g., rule curves; Loucks and Van Beek 2017). Until today, not much effort was made to target DPS approaches more specifically to dam operators and their traditional way of operation. Giuliani et al. (2014b) developed a framework for operators to redesign operating rules for a many-objective reservoir, including uncertainty in the hydro-climatic variables and involving visual analytics, with the aim to get closer to the perspective of dam operators. The term "many-objective" describes multi-objective problems with more than three objectives, which are often faced by operators in the real world (Purshouse and Fleming 2003; Giuliani et al. 2014a). However, in the mentioned study, policies remain to be expressed by RBF approximators and are therefore not related to the rule-based control that operators are used to. Herman and Giuliani (2016) presented an approach to the design of easily interpretable policies as a tree structure of logical rules for a single reservoir, with the motivation to help operators adjust operating rules to possible future scenarios. Still, there is a clear need to do more in this direction.

The main motivation for the current study is the existing problem of insufficient physical representation in the parameterized functions in DPS. Therefore, the release policies have been expressed in terms of operating rules that resemble the actual decision-making process in the daily operation of reservoirs. The operating rules include parameters (decision variables) that are optimized with respect to the operating objectives. For instance, such a rule could be lowering the level in a dam during the wet season to avoid spills; in this case the decision variable is the target level in the dam. These "traditional" policies are closer to real practice and are directly interpretable, and this makes them potentially more acceptable in the mindset of traditional reservoir operators. There is, of course, a tradeoff: the adopted policies are based on traditional operating rules, so even if they are optimized, they may be worse than those using DPS approaches that employ more flexible mathematical functions. For single reservoirs, such physically based operating rules were optimized (also taking into account multiple objectives) by a number of studies (e.g., Galelli et al. 2014). However, the complexity of this task increases rapidly when networks consisting of several reservoirs are considered.

The proposed version of DPS is applied to the hydropower reservoir network in the Nechí River in Colombia, consisting of five reservoirs. This system was found to be a good case study for the proposed approach, since all five dams of the network are operated by the one company, which facilitates the implementation of a common operational strategy. In the following chapters, operating rules are developed and optimized for this particular system. As such, they are not directly transferrable to other reservoirs networks, but the rules can serve as the basis to formulate similar operating rules for other systems, only requiring adjustments to the given physical characteristics.

## Case Study and Problem Description

The Nechí River is located in northwest Colombia between the Cauca and the Magdalena Rivers (Fig. 1a). The dry season is from December through March and the wet season is from from April to mid-November, with strong rainfall events that last from several days to weeks. Annual precipitation is around 3,500 mm; however, the uneven distribution of annual precipitation is reflected in high differences of overall system inflow over the year (Fig. 2). This represents a big challenge in terms of flood and drought management.

Flooding occurs almost annually in the downstream part of the catchment and causes casualties and economic damage, especially in the cities of El Bagre and Zaragoza. Apart from that, the Nechí is lined with legal and illegal gold mines that heavily pollute downstream water bodies when they are flooded. On the other hand, the hydrologic variation is crucial to support a variety of ecological processes and ecosystem services in the lower Nechi River floodplains, including critical food supplies based on native fisheries.

In the dry season, water levels in the reservoirs drop drastically, a problem that is often exacerbated by the occurrence of El Niño that further reduces system inflows (Fig. 2). This jeopardizes energy security in Colombia, whose energy mix relies on hydropower for around 70% of its production (Procolombia 2015). It is therefore critical to operate hydropower dams in a fashion that enables sufficiently high energy production to cover national demand also in the dry season.

Under current operations, the network is operated to maximize the revenue from electricity generation. Additional interests such as reduction of flood hazard or hydrologic flow alteration were not accounted for in the original design of the structures and their operating rules. They are — if at all — included only as rigid constraints, such as dam security buffers or minimum environmental flows (MEFs); the latter are prescribed by the National Authority of Environmental Licenses (ANLA). Integrating all involved interests could lead to release policies that result in better overall performances of the system, and expressing these policies in the physically oriented

"language" of dam operators may be an approach to overcome policy myopia.

In the upstream part of the catchment, a network of five reservoirs and four hydropower facilities with an overall installed power of 1,883 MW provides around 15% of Colombia's electricity (Procolombia 2015).

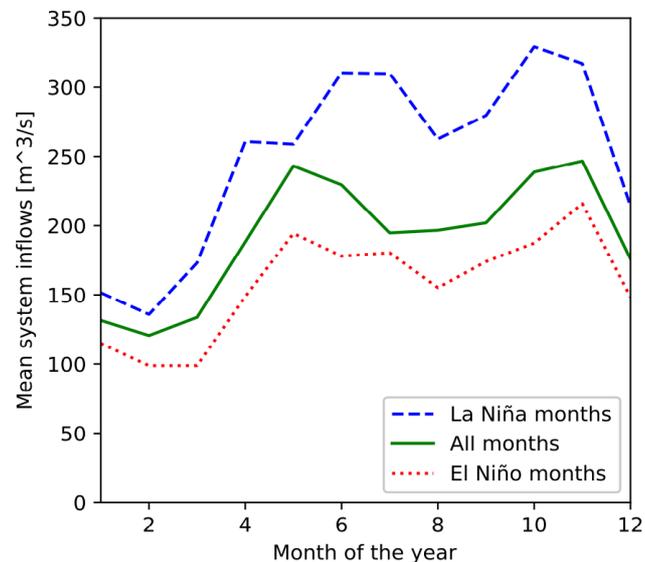

Figure 2. Mean monthly system inflows (2002–2015). The classification of months is based on the Oscillation Niño Index signal (ONI; El Niño: ONI > 0.5; La Niña: ONI < −0.5).

The network has two branches that meet just upstream of Porce III, the most downstream reservoir of the cascade (Fig. 1b). The northern branch starts with flow diversions from the upper Nechí. These diversions are — along with releases from Miraflores, which merely serves as a drought buffer — further diverted through an underground channel into Troneras. From Troneras, a chain of three power plants called Guatron is supplied with water, which releases just upstream of Porce III. The biggest reservoir of the system, Riogrande II, is the starting point of the southern branch. Most of its water is diverted to the La Tasajera power plant, making use of more than 900 m of hydraulic head. Through the river outlet of the dam, only MEFs and spills are released. All releases eventually flow into River Porce, which feeds the Porce II dam. After passing the turbines of Porce II, the water reaches the confluence with the northern branch before flowing into Porce III.

## Operational Model

The analysis of the reservoir network resulted in a schematization of the system, which was the base for the operational model (Fig. 3). The daily inflow time series are represented by grey ellipses and represent runoff from the natural drainage areas of the dams (except $IN_{Diversions}$ and $IN_{Concepcion}$ that are diverted flows without controllable inflow structures). The daily inflow time series are freely available for 2002-2015 from XM Compañía de Expertos en Mercados (2017), except $IN_{Porce3}$, which is available from 2011, the year when Porce III was put into operation. Therefore, $IN_{Porce3}$ has been extended to 2002-2015 with a validated data-driven model of high accuracy (RMSE 18.4 m$^3$/s, correlation coefficient 0.9869) that has been created using the Weka Workbench software (Frank et al., 2016) with the M5 Rules classifier. For more information on the data-driven model and its accuracy, see Appendix S1. After this extension, the available data and thus the modeling horizon encompass 14 years. Flow lag times between structures have been neglected, as the longest is in the range of only six hours (La Tasajera – Porce II).

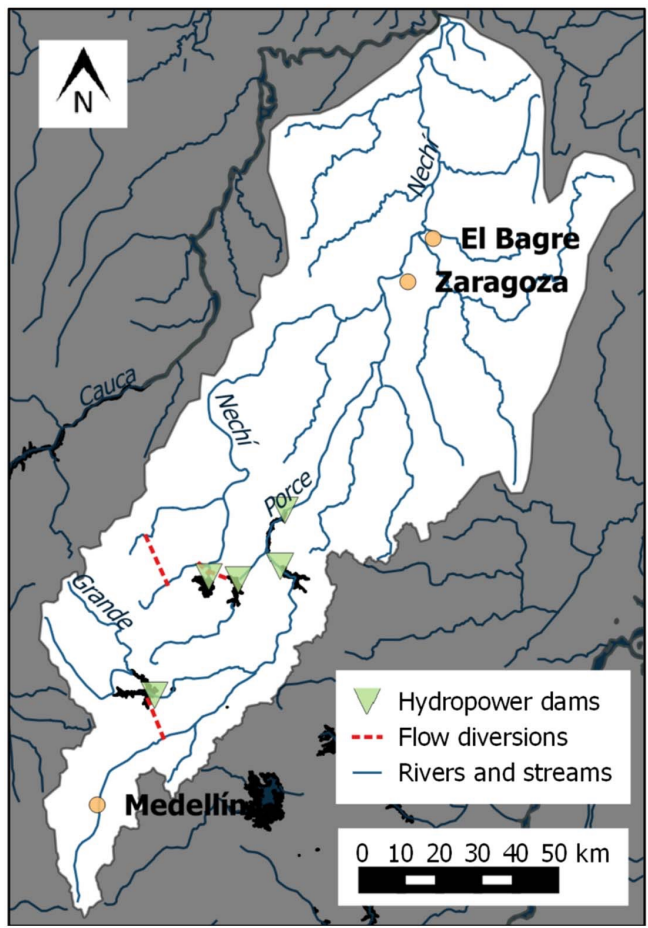

(a)

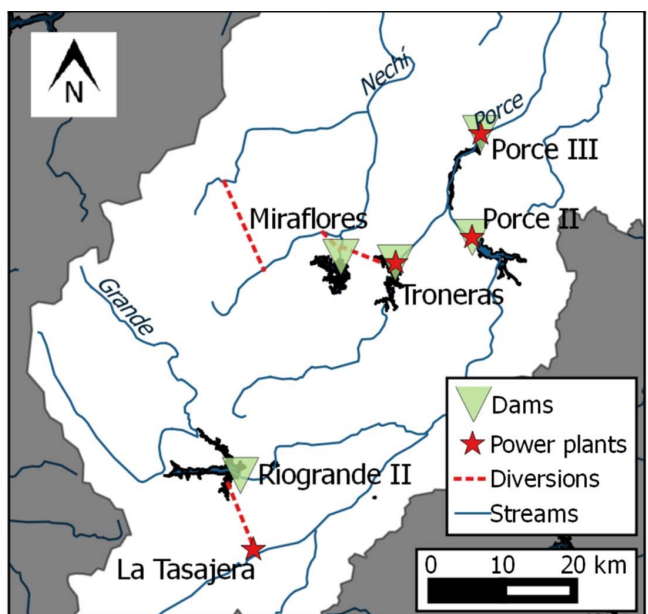

(b)

Figure 1. Map of: (a) Nechí catchment; and (b) the hydropower dam cascade.

Cooperation with the operating company Empresas Públicas de Medellín (EPM), the Colombian Institute of Hydrology (IDEAM), and the National Authority for Environmental Licenses (ANLA) enabled the authors to make a reasonably detailed representation of the physical processes and engineering systems involved. On the basis of the ReservoirSimulator (Angarita et al., 2013, 2018; Ritter, 2016), the system model includes volume-elevation-curves of the reservoirs, efficiency curves of turbines, evaporation losses, MEF regulations, inter alia. At every time step (i.e. day), the model computes the water balance for each of the five reservoirs, going from upstream to downstream to account for inflows originating from upstream dams:

$$s_{t+1} = s_t + IN_t - E_t - q_{MEF,t} + q_{in,t} - q_{turb,t} - q_{spill,t} \quad (1)$$

where $s_{t+1}$ (subject to $s_{t+1} \geq 0$) and $s_t$ stand for the storage in the reservoir at the following ($t + 1$) and the current time step ($t$), respectively. For natural inflows $IN_t$ (Fig. 3), a perfect foresight of one time step has been assumed, which seemed reasonable considering that the reservoir capacities are large compared to daily peak system inflows (i.e. for the most critical reservoir Porce II, observed inflows never exceeded 33% of the downstream reservoir's capacity). Evaporation loss $E_t$ is estimated on the basis of the current surface area of the reservoir and the month of the year. Potential MEF requirements between the dam and the turbine outlet (Porce II and Porce III) or into historical river outlets (Riogrande II and Miraflores) are represented by $q_{MEF,t}$; inflows from upstream dams are $q_{in,t}$; water released through the turbines is $q_{turb,t}$, and via the spillway $q_{spill,t}$. For every time step and reservoir, the water balance Eq. 1 is solved twice: firstly, before the release decision ("preliminary"); secondly, after the release decision ("final"):

1. The preliminary water balance is used to obtain an estimation of the available water volume in the dam at the current time step before release decisions are made. The inflow from upstream dams $q_{in,t}$ is still uncertain and thus needs to be approximated. This is done by multiplying the filling percentages of upstream dams by the maximum discharge capacity of their turbines. $q_{turb}$ and $q_{spill}$ are equal to zero.
2. The final water balance after the release decisions is necessary to determine the exact storage of the dam for the next time step. The inflow from upstream dams $q_{turb}$ is here determined by the operating rules in the decision-making component (see following section), and $q_{spill} = (s_{t+1} - s_{max}) \geq 0$, where $s_{max}$ is the capacity of the reservoir.

**Decision-Making Component**

Every time step (i.e. every day), the decision on the release $q_{turb}$ (Eq. 1) for each of the five dams has to be made. As described in the introduction, the goal of this study is to do this is in a way that resembles the actual decision-making process of dam operators, instead of optimizing black-box functions. This requires information about potential inputs and strategies that could be useful to control a network of reservoirs. Operating companies keep this information under tight wraps since it could make their energy generation predictable by competitors on the energy market, potentially resulting in financial losses. Therefore, this information had to be approximated with the help of experts in the field of water management in Colombia from IDEAM and the Escuela Colombiana de Ingeniería. The brainstorming process resulted in a list of potentially useful inputs to control the system (Table 1), and in useful insights on the influence of the Colombian energy market on daily decisions of dam operators.

Numerous studies show that cooperation between operators of cascaded hydropower dams leads to an increase of overall benefits (e.g. Yu et al., 2019; Wu et al., 2016; Chen et al., 2013). In the Nechí catchment, all reservoirs are operated by one company (EPM), and this fact considerably facilitates cooperation. To reduce the number of decision variables, decision-making at each time step has been divided into two parts, following the main idea of the widely used aggregation-decomposition approach (e.g. Liu et al., 2011; Tan et al., 2017). First, the system states of all reservoirs are taken into account to make a first "global" decision for the reservoir network. Then, this decision is decomposed into individual "local" decisions for the individual reservoirs. Applied to the decision-making process in the hydropower cascade of the current study, this means the following: Firstly, the global energy generation target for the entire lumped cascade and the current time step is computed (DM component – Part 1). Secondly, this global target is allocated to the individual dams and power plants (DM component – Part 2).

In this paper, energy generation targets are expressed as unitless values in the range [0; 1], representing the fraction of the installed power of the turbines. For instance, a global generation target ($GT_t$) of 0.5 means that the goal of the current time step is to generate 50% of overall installed power (in our case 0.5 x 1883 MW = 941.5 MW). Analogously, the local energy target of dam $i$ ($LT_{i,t}$) is expressed as the fraction of installed power of its turbines.

Fig. 4 shows the general setting of the DM component and its connection to the operational model. The main outputs of the model are releases and generated power of the dams at every time step t. The individual power outputs are summed at every time step to compute the overall generated energy of the system. Over the control horizon, this returns a time series of overall power generation. The discharge values of the most downstream dam Porce III over the control horizon represent the output hydrograph of the system. Only these two time series are later evaluated by the objective functions.

The following sections explain the process presented in Fig. 4 in detail. The computation of the energy targets and the resulting releases at each time step is done depending on the system states and 6 operating rules, with a total of 12 decision variables (Table 2) that are based on the previously identified inputs (Table 1).

Table 1. Potentially useful inputs for system control, with indications if they are used for calculation of the global target (i.e., daily energy generation target for all dams combined), its allocation to the individual dams, or release from Miraflores to Troneras.

| Input | Global target | Allocation | Miraflores |
|---|---|---|---|
| Filling percentages of individual dams (system state) | X | X | X |
| Position of individual dams in network (upstr./dwnstr.) | X | X | — |
| Energy market price | X | — | — |
| Persistence in monthly inflows (seasonal inflow forecast) | X | — | — |
| El Niño/La Niña influence | X | — | — |

### Decision-Making Component Part 1: Calculation of the Global Energy Generation Target ($GT_t$)

The global energy generation target [0; 1] at the current time step is calculated as a linear combination of the inputs in Table 1:

$$GT_t = v_1 \cdot filling_t + v_2 \cdot price_t + v_3 \cdot inflows_t + v_4 \cdot enso_t \quad (2)$$

The weights of the four factors representing the inputs have been defined as decision variables $v_1$ to $v_4$ (all [0; 1], subject to $\sum v = 1$). All four factors are normalized to [0; 1]; this is required to have the possibility to combine very different types of information (such as energy price and filling percentages) to one value that represents the global generation target. The calculations of the four factors are explained in the following:

1. Weighted overall filling percentage: The first factor $filling_t$ in Eq. 2 describes the system state (aggregate storage), taking into account that water in upstream dams is more "valuable" than water in the downstream dams, as it passes more turbines on its way through the system. It combines the first two inputs of Table 1 in one equation, giving higher weight to water in upstream dams:

$$filling_t = \frac{\sum_{i=1}^{n}\left(\frac{InstP_{dwnstr,i}}{\sum_{h=1}^{n}InstP_{dwnstr,h}} \cdot s_{i,t}\right)}{\sum_{i=1}^{n}\left(\frac{InstP_{dwnstr,i}}{\sum_{h=1}^{n}InstP_{dwnstr,h}} \cdot s_{i,\max}\right)} \quad (3)$$

where $n$ is the number of reservoirs in the network; $s_{i,t}$ is the storage of reservoir $i$ before release (calculated in the preliminary water balance, see Operational Model) and $s_{i,max}$ is the maximum storage; $InstP_{dwnstr,i}$ and $InstP_{dwnstr,h}$ are the sums of installed powers in all power plants downstream of dam $i$ and $h$, respectively (Table 3). For instance, in the case of the Nechí, the network has two branches. Only the installed powers downstream of reservoir $i$ (i.e. not the ones in the other branch) play into $InstP_{dwnstr,i}$. Note again that Miraflores does not have any turbines ($InstP_M = 0$). By default, $filling_t$ is within [0; 1] and thus does not require normalization.

2. Energy market price: Higher energy prices give incentives to generate more electricity. When water availability is low (i.e. in a dry season) energy price is high. The factor $price_t$ in Eq. 2 accounts for this with a cyclostationary pattern of mean monthly anomaly fraction of the annual price, based on daily averages of Colombian energy prices (2002-2015):

$$price_{month,j} = \frac{1}{(a_{end} - a_{start} + 1)} \cdot \sum_{a_{start}}^{a_{end}} \frac{\overline{price_{month,j,a}}}{\overline{price_{year,a}}} \quad (4)$$

where $a$ are the start and end years of the data period, $\overline{price_{month,j,a}}$ is the average price during month $j$ of year $a$ and $\overline{price_{year,a}}$ is the average price of year $a$. Since all four factors in Eq. 2 have to be in the range [0; 1], the resulting 12 values of $price_{month}$ (one for each month of the year) have also been normalized to [0; 1]. At every time step, the value of $price_t$ (Eq. 2) is set corresponding to the current month of the year.

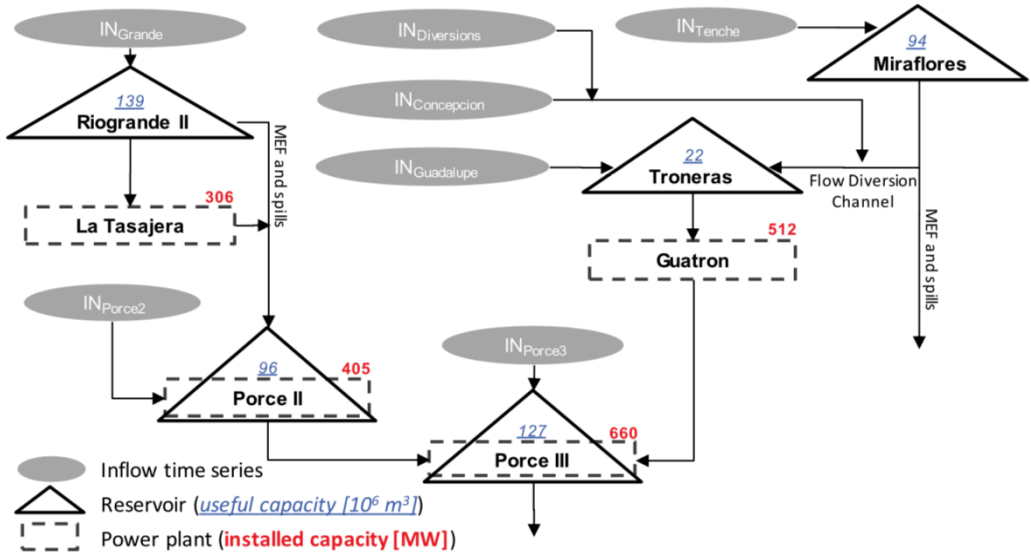

*Figure 3. Schematization of the reservoir network and its inflow data time series. Note that the power houses of Porce II and Porce III are not inside the dam structure, but several kilometers downstream. This is important for MEF requirements between the dam structures and the turbine outlets.*

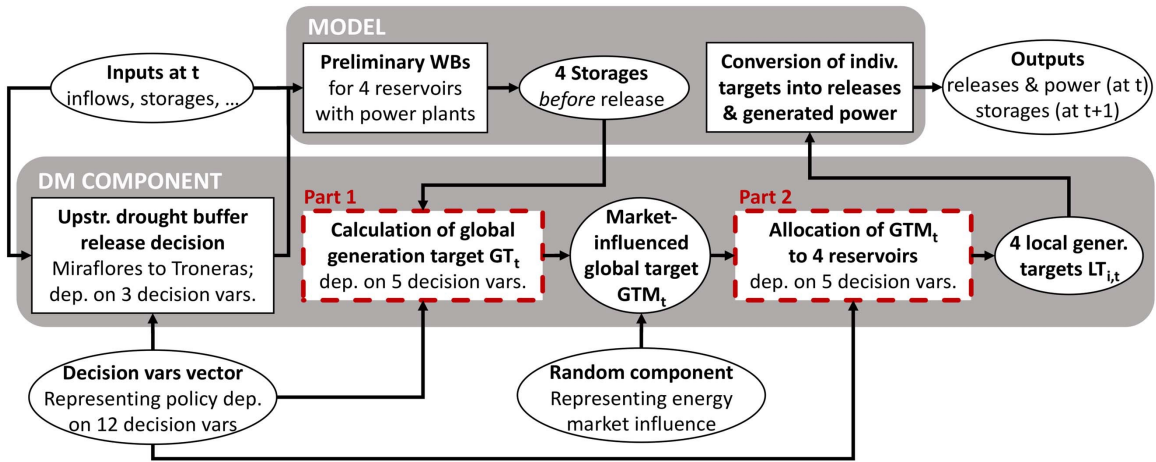

*Figure 4. Schematization of the system model and the decision-making component. The presented workflow is repeated at every time step (t) over the control horizon. The main results are the time series of overall power generation and discharge downstream of the cascade. These two results are later evaluated by the objective functions. WB stands for water balance. The decision variable $y_{drywet}$, is used in "Upstream drought buffer release decision" and "Part 2 of the DM component."*

Table 2. Summary of the 6 operating rules and the 12 decision variables involved.

| Purpose | Rule | Decision variables | Short description | Range |
|---|---|---|---|---|
| G | Global generation target | $p$ | Weight decline toward future months (persistence in monthly inflows) | [0; 15] |
| | | $v_1$ | Weight of weighted filling percentage (system state) | [0; 1] |
| | | $v_2$ | Weight of energy market price | [0; 1] |
| | | $v_3$ | Weight of persistence in monthly inflows | [0; 1] |
| | | $v_4$ | Weight of El Niño/La Niña influence | [0; 1] |
| A | Full reservoir | — | — | — |
| | Weight of low flow alteration rule (Porce III) | $m_{P3}$ | Weight of low flow alteration rule Porce III | [0; 2] |
| | Flood buffers (Porce II and Porce III) | $b_{P2}$ | Target filling percentage of Porce II in wet season | [0; 1] |
| | | $b_{P3}$ | Target filling percentage of Porce III in wet season | [0; 1] |
| | | $y_{drywet}$ | Inflow threshold between dry and wet season | [0; 250] |
| | Allocation of remaining target | $a_{alloc}$ | Weight in allocation rule between filling-% and location of dams | [0; 1] |
| M | Release from Miraflores | $r_M$ | Parameter in release function of Miraflores | [0; 23.6] |
| | | $f_T$ | Target filling percentage of Troneras in dry season | [0; 1] |
| | | $y_{drywet}$ | Inflow threshold between dry and wet season | [0; 250] |

Note: The first column indicates the purpose of the rule. G stands for global generation target, A stands for allocation of the global target to individual dams and M for the release from Miraflores to Troneras. The decision variable $y_{drywet}$ has a double function in A and M. The purpose of the full reservoir rule is to run turbines at full speed if the reservoir is at its technical capacity. This does not require optimization and therefore has no decision variables assigned.

Table 3. Calculation of Inst$_{Pdwnstr}$, the combined installed power of all turbines downstream of each dam i.

| Reservoir | InstP (MW) | InstP$_{dwnstr}$ (MW) |
|---|---|---|
| Porce III | 660 | $InstP_{P3} = 660$ |
| Porce II | 405 | $InstP_{P2} + InstP_{P3} = 1,065$ |
| Riogrande II | 306 | $InstP_{R2} + InstP_{P2} + InstP_{P3} = 1,371$ |
| Troneras (Guatron) | 512 | $InstP_G + InstP_{P3} = 1,172$ |
| Miraflores | 0 | $InstP_M + InstP_G + InstP_{P3} = 1,172$ |

Note: Required for giving higher weight to water in upstream dams in the calculation of the component $filling_t$.

3. Persistence in monthly inflows: When assessing inflows, the following considerations have been taken into account. As there was no seasonal rainfall forecast available for the given area and modelling horizon (2002-2015), a cyclostationary pattern of overall system inflows has been derived from the inflow time series by creating a function of mean monthly system inflows over the course of the year ($\overline{IN}_{total,month}$; see Fig. 2). Based on comparisons between reservoir capacities and mean inflows, it has been considered reasonable to assume that the inflows of the current month and of up to three months ahead could be potentially relevant for operation. To determine the relative weights of the current and three future months, the approach of the Rank Exponent Weight Method has been employed (e.g. Roszkowska, 2013), and adjusted to our problem:

$$w_u = \frac{(m-u+1)^p}{\sum_{u=1}^{m}(m-u+1)^p} \quad (5)$$

where $m$ is the number of considered months (in our case $m = 4$) and $u$ [1; $m$] is the integer month number (current month: $u = 1$, next month: $u = 2$, …). This formula computes weights $w_u$ for the four months, assigning the highest weight for the current month, and progressively declining weights for the following months. The weight decline has been made subject to optimization by means of the decision variable $p$. The selected range of $p$ is [0; 15], allowing for equal weights ($p = 0 \rightarrow w_{1-4} = 0.25$) and nearly full weight on the current month ($p = 15 \rightarrow w_1 = 0.99$). For each month $j$ of the year, the weights are applied to the current and the three following mean monthly inflows, and the weighted values are summed:

$$inflows_{month,j} = \sum_{u=1}^{m} w_u \cdot \overline{IN}_{total,month,j+u-1} \quad (6)$$

Finally, these 12 weighted sums are again normalized to obtain for each month $j$ the value of $inflows_{month,j}$ [0; 1]. At every time step, the value of $inflows_t$ (Eq. 2) is set corresponding to the current month of the year.

An alternative to the presented way of providing the months' weights (Eq. 5) would be to assign a separate decision variable for each of the months. However, this would have increased the number of decision variables and thus the computational complexity, and therefore this option has been discarded.

4. El Niño/La Niña influence: The fourth factor in Eq. 2 takes into account the effects of El Niño and La Niña phenomena. These are commonly described by the El Niño Southern Oscillation (ENSO). The primary indicator to monitor ENSO is the Oscillation Niño Index signal (ONI), which is provided in three-month-moving-average values of temperature anomalies in the Pacific Ocean (NOAA, 2015). In the modeling horizon (2002-2015) ONI oscillated within [-1.4; 2.3] (i.e. El Niño: ONI > 0.5; La Niña: ONI < -0.5). In Colombia, El Niño reduces rainfall, while La Niña brings more precipitation than usual (see Fig. 2 for the effect on system inflows), provoking lower and higher generation targets in hydropower dams, respectively (Poveda et al., 2001). Therefore, ONI has been inverted and then also normalized, so the factor $enso_t$ is in the range [0; 1]. Simply put, this means that a value of $enso_t = 0$ corresponds to ONI = 2.3, and $enso_t = 1$ to ONI = -1.4. For potential future ONI values exceeding the observed, $enso_t$ is constrained to [0; 1].

There is yet another factor to be taken into account in relation to the global generation target (Eq. 2): the influence of the energy market, which strongly depends on the (highly confidential) business strategy of the operating company. In Colombia, generating companies have two options to sell their energy: using long-term contracts, or on the free market. In practice, the business strategies usually combine these two options (Cuadros and Ortega, 2012). Offers to the free market can be rejected, even only for individual hours of the day. In contrast, when energy security is at stake, the market can give additional incentives to generate (Cramton and Stoft, 2007; de Castro, 2016). This makes it difficult to predict market influence on daily generation targets.

To make the optimized policies more robust regarding these external influences, a random factor has been introduced to transform $GT_t$ into the generation target under market influence $GTM_t$. As the market can have an increasing and a decreasing influence on the generation target, $GTM_t$ has been chosen to be a random variable normally distributed around $GT_t$. A standard deviation of $GT_t/20$ has been assumed, based on a rough estimation from the consulted Colombian hydropower experts (see Decision-making Component – General description): around 5% of the time, offers are rejected by the market. This representation of the market influence is a very rough estimation and could be certainly improved if data availability allowed for supporting the choice of the probability distribution function and its parametrization. In this study, the required data was unfortunately not available and thus had to be approximated by consultations with experts. A system operator may dispose of more information and data and adjust this component accordingly.

$$GTM_t = GT_t + N\left(0, \frac{GT_t}{20}\right) \in [0; 1] \quad (7)$$

For the current time step, this market-influenced target $GTM_t$ is now a fixed value [0; 1]. To obtain its value in [MW], it is multiplied by the overall installed power of the lumped cascade (in this study 1883 MW).

### *DM Component Part 2: Allocation of $GTM_t$ to the individual power plants*

After calculating $GTM_t$, a set of operating rules has to be applied to allocate it to the individual four dams with power plants (Miraflores does not have any turbines; its role will be discussed later). First, three operating rules that are based on constraints allocate a first portion of $GTM_t$. Then the remaining portion is allocated based on filling percentages and relative positions of the individual reservoirs. The four operating rules that take charge of the allocation are described in the following:

1. Full reservoir rule (applies to all reservoirs): It is always better to release as much water as possible through the turbines, instead of the spillway or other conduits (which happens if a reservoir gets full). To reduce these spillages, a constraint is applied to all reservoirs: If the preliminary water balance (before release; see Operational Model) in reservoir $i$ resulted in storage greater or equal to its capacity ($s_{t+1} \geq s_{max}$), the local generation target $LT_{i,t}$ is set to the maximum ($LT_{i,t} = 1$).

2. Flow alteration reduction rule (applies to Porce III only): Porce III is the most downstream dam and thus the only one able to effectively mitigate hydrologic flow alteration downstream of the cascade. Flow alteration is an indicator that is commonly used to describe the environmental impact downstream of dams, e.g. by Vogel et al. (2007). When dams are not generating energy (which happens in practice very frequently, mainly due to energy market dynamics), they usually release only very little water. For instance, when the turbines of Porce III are switched off, only 2 m³/s are released to satisfy MEF regulations. This value does not compare to natural discharges (see Fig. 2) and this has diverse impacts on the downstream ecosystem. To mitigate this alteration in the low flows, a minimum turbine discharge constraint $Q_{P3min,month,j}$ [m³/s] for Porce III is proposed. Unfortunately, there is no information available on critical flow thresholds, below which significant environmental impacts occur. Therefore, this information was approximated based on the natural hydrograph, which allowed for proposing a minimum target flow for each month $j$ of the year:

$$Q_{P3min,month,j} = m_{P3} \cdot MEQ_{month,j} \quad (8)$$

where $MEQ_{month,j}$ [m³/s] is the minimum discharge ever observed in the natural hydrograph in month $j$ of the year. It is multiplied by the decision variable $m_{P3}$ [0; 2] to give a relative weight to the rule. For instance, $m_{P3} = 0$ would disable the rule, while $m_{P3} = 2$ would mean that the minimum release constraint for Porce III is twice the minimum discharge ever observed in the current month of the natural hydrograph. At each time step, the value of $Q_{P3min,t}$ is chosen from $Q_{P3min,month,j}$ according to the current month of the year. By applying the turbine equation, $Q_{P3min,t}$ [m³/s] is converted into the corresponding minimum target generation $LT_{P3,min,t}$ [0; 1] of Porce III:

$$LT_{P3,min,t} = \frac{Q_{P3min,t} \cdot H_t \cdot (1-\xi) \cdot \gamma \cdot \eta_t}{InstP_{P3}} \quad (9)$$

where $InstP_{P3}$ the installed power [MW] and $H_t$ the current hydraulic head of Porce III; $\xi$ is the pipe loss, $\gamma$ is the specific weight of water and $\eta_t$ is the turbine efficiency for the current discharge and hydraulic head.

3. Flood buffer rules (apply to Porce II and Porce III only): In test runs of the model, it became clear that the capacity of Porce III alone is not sufficient for an effective flood hazard reduction, especially when wet season coincides with the occurrence of La Niña (Fig. 2). As described in the Case Study and Problem Description, rainfall events in wet season may last for several weeks, and keeping an empty buffer capacity only in Porce III does often not result in a significant lowering of the hydrograph peak. Therefore, also Porce II has been included in flood hazard reduction strategy. A conditional rule is proposed that lowers the water levels in these two downstream dams during the wet season, in order to create buffer capacities. The following equations describe the rule for Porce III, but it applies analogously to Porce II (exchange subscript P3 by P2):

$$if \quad \frac{1}{8}\sum_{t-7}^{t} IN_{total,t} > y_{drywet} \quad (10)$$

$$and \quad \frac{s_{P3,t}}{s_{P3,max}} > b_{P3} \quad (11)$$

$$then \quad LT_{P3,t} = 1 \quad (12)$$

where $IN_{total,t}$ [m³/s] is the total system inflow and $s_{P3,t}$ is the storage of Porce III at time step $t$ (before release; see Operational Model). The maximum storage is $s_{P3,max}$. To decide when flood buffer rules are triggered, the decision variable $y_{drywet}$ [0; 250 m³/s] has been introduced as a discharge threshold. The average of system inflows of the seven days prior to the current time step and the current time step has been defined as the criterion to distinguish between the dry and the wet season. If this average is higher than the threshold $y_{drywet}$ (Eq. 10), flood buffer rules are triggered. The decision variable $b_{P3}$ [0; 1] has been assigned (analogously $b_{P2}$ [0; 1] for Porce II) and represents the target filling percentage of Porce III during the wet season. If the ratio of the storage (before release) at the current time step $s_{P3,t}$ to the maximum capacity $s_{P3,max}$ is higher than this target filling percentage $b_{P3}$ (Eq. 11), the local target generation $LT_{P3,t}$ is set to the maximum (Eq. 12). The maximum local target results in maximum turbine release and thus aims at lowering the water level for a larger buffer capacity in the dam.

4. Allocation of the remaining target (applies to all reservoirs with power plants)**:** After applying the three rules described above, a portion of $GTM_t$ may be allocated, reducing it to $GTM_{remaining,t}$. This remaining portion is now further allocated to the power plants. This is done by an iterative procedure that allocates to all power plants with remaining turbine capacity ($LT_t < 1$) and water availability ($s_t > 0$). The iterative procedure stops when $GTM_{remaining,t}$ is reached (i.e. global target fully allocated), or when water availability does not allow for further allocation. Two inputs of Table 1 have been considered: the filling percentages, and the relative position of each dam in the network. The idea of including the position is to generally assign higher targets to downstream dams to lower their water levels (compared to upstream dams) for flood protection. $R_k$, the fraction of $GTM_{remaining,t}$ that is allocated to dam $k$, is computed as:

$$R_k = a_{alloc} \cdot \frac{fill_{k,t}}{\sum_{i=1}^{N} fill_{i,t}} + (1 - a_{alloc}) \cdot \frac{1 - \frac{InstP_{dwnstr,k}}{\sum_{i=1}^{N} InstP_{dwnstr,i}}}{\sum_{i=1}^{N} \left(1 - \frac{InstP_{dwnstr,k}}{\sum_{i=1}^{N} InstP_{dwnstr,i}}\right)} \quad (13)$$

where the first term represents the filling percentage [0; 1] of dam $k$ in relation to the sum of filling percentages of all $N$ dams with power plants (in our case $N = 4$, as Miraflores does not have turbines). As opposed to the weighted filling percentage $filling_t$ in Eq. 3, these are mere filling percentages and do not take the cascade configuration into account. The second term of Eq. 13 includes information on whether dam $k$ is upstream or downstream in relation to the other dams. For the values of the installed power downstream of dams $InstP_{dwnstr}$, refer again to Table 3. The decision variable $a_{alloc}$ [0; 1] defines the weight distribution of the two inputs in the allocation procedure. In case a power plant is allocated with an amount of energy that exceeds its capacity ($LT_t > 1$), the excess is allocated to the remaining power plants by means of a further iteration of Eq. 13.

The four local generation targets $LT_{i,t}$ [0; 1] are converted into absolute power targets [MW] by multiplying them by the installed capacity of their power plant. Each of the four dams contains several turbines (in the given system, up to six). In order to hold back as much water as possible while reaching $LT_{i,t}$, the operational model applies turbine efficiency curves to identify for each dam the optimal number of activated turbines at the current time step. The results for each dam are (a) the turbine discharge $q_{turb,t}$ in the final water balance (Eq. 1), and (b) the value of generated power $P_t$ at the current time step.

As explained earlier, Miraflores serves as drought buffer for the system, which already implies a difference in operation between seasons. The decision variable $y_{drywet}$, which was earlier introduced for triggering flood buffer rules, is used again here to trigger the dry season operation of Miraflores:

$$if \; \frac{1}{8}\sum_{t-7}^{t} IN_{total,t} < y_{drywet} \tag{14}$$

$$and \; \frac{s_{T,t}}{s_{T,max}} < f_T \tag{15}$$

$$then \; Q_{Miraflores,t} = Q_{DVChannel,max} - IN_{Diversions,t} - IN_{Concepcion,t} \tag{16}$$

where $Q_{DVChannel,max} = 23.6$ m³/s is the maximum discharge capacity of the diversion channel to the reservoir Troneras. The diversion channel is first filled with the inflow time series $IN_{Diversions,t}$ and $IN_{Concepcion,t}$ (see Fig. 3). The remaining discharge capacity is available for release from Miraflores. The conditional rule does the following: During the dry season (Eq. 14), release from Miraflores to Troneras is equal to the remaining discharge capacity of the diversion channel (Eq. 16), if the filling percentage in the downstream reservoir Troneras drops below the decision variable $f_T$ [0; 1] (Eq. 15). In other words, $f_T$ is the target filling percentage of Troneras during the dry season. In the wet season, release from Miraflores is controlled by two linear functions (presented as planes in Fig. 5). The two functions depend on the decision variable $r_M$ [0; 23.6 m³/s]. The release is always subject to the remaining discharge capacity in the diversion channel.

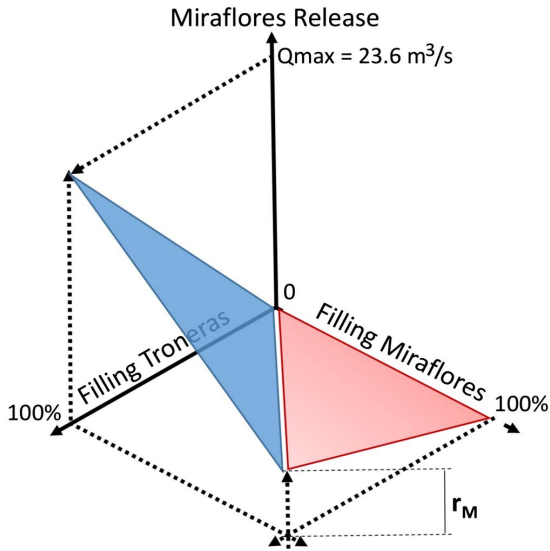

Figure 5. Control functions of release from the most upstream dam, Miraflores, through the diversion channel toward Troneras in the wet season.

## Objective functions

Four objectives functions (OFs) have been considered; they evaluate the output time series (2002-2015) of the operational model with the described operating rules: (a) the time series of overall generated power of the lumped cascade $P_{total}$ (evaluated by OF1 and OF2), and (b) the hydrograph downstream of the cascade $Q_{total}$ (i.e. downstream of Porce III; evaluated by OF3 and OF4). Please note again that the energy generation targets $GT$, $GTM$ and $LT_i$ are not directly evaluated. Their only role is to connect the inputs (Table 1) to the power generation and release time series.

OF1. Average generated energy: The average power generated by the entire system is to be maximized and is evaluated by the following objective function:

$$J_{max}^{avgE} = \frac{1}{T}\sum_{t=1}^{T} P_{total,t} \tag{17}$$

where $T$ is the number of time steps and $P_{total,t}$ is the generated power of the system at time step $t$.

OF2. Firm energy: To safeguard energy security in the country, the government in Colombia imposes a minimum power generation requirement on energy providers in water scarcity conditions. In the dry season, hydropower stations often struggle to generate the so-called firm energy that they are legally obliged to provide. By definition, firm energy is the monthly average power that can statistically be exceeded in 95% of the months (Fig. 6). Therefore, firm energy is to be maximized and is evaluated as the 5th percentile of all monthly averages of power generation $\overline{P_{total,month}}$ [MW] with M being the number of simulated months:

$$J_{max}^{firmE} = PCTL_5(\overline{P_{total,month,1}}, \ldots, \overline{P_{total,month,M}}) \tag{18}$$

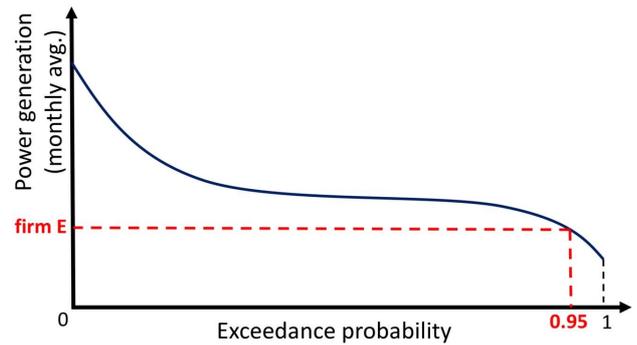

Figure 6. Definition of firm energy as the 5th percentile in the duration curve of monthly averages of energy generation.

OF3. Flood hazard: Flood hazard is evaluated by identifying yearly discharge peaks from the hydrograph downstream of the cascade $Q_{total}$. The yearly maxima are sorted in descending order and integer ranks $r_y$ [1; Y] are assigned (Y is the number of simulated years). For instance, $r_y = 1$ expresses that the corresponding discharge is exceeded in one of $Y$ years. Then, the return period (*RT*) for each peak discharge is calculated by computing *RT* = *Y* / *r*. The result is a curve of RT versus corresponding peak discharge (Fig. 7). Analogous to the computation of Expected Annual Damage (EAD), which is a common practice in flood risk management (e.g. Arnell, 1989), the (expected annual) flood hazard is the area under the annual peak flows curve. This area is additionally constrained based on the following idea: for releases up to the maximum turbine capacity of Porce III ($Q_{total} \leq Q_{P3,max}$), flood hazard was assumed to be zero. In other words, yearly peaks only play into the evaluation of flood hazard, when spills occurred in the corresponding year. The objective function minimizes the resulting (constrained) area under the annual peak flows curve and above the maximum turbine capacity, which has the unit [year · m$^3$/s]:

$$J_{\min}^{flood} = \int_{RT(Q=Q_{P3,\max})}^{Y} Q(RT) - Q_{P3,\max} dRT \quad (19)$$

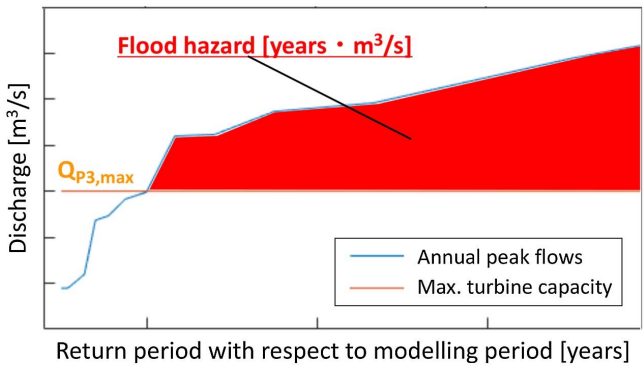

*Figure 7. Calculation of (expected annual) flood hazard analogous to the concept of expected annual damage (EAD).*

OF4. Flow regime alteration: For the evaluation of this objective, the ELOHA dashboard of The Nature Conservancy has been employed, a tool for evaluating flow regime alteration. Following the approach of Vogel et al. (2007), flow alteration has been quantified by comparing flow-duration-curves (FDC) of impaired hydrograph and a natural reference hydrograph. The natural reference has been approximated by summing the natural inflow time series of the system. This implies all inflow time series except **IN$_{Diversions}$**, **IN$_{Concepcion}$**, and **IN$_{Tenche}$** (see Fig. 3).

To identify flow alterations, 12 mean monthly FDCs are computed for the modelled hydrograph downstream of the cascade $Q_{total}$. Plotting for each month of the year the impaired FDC against the mean monthly FDC of the natural reference hydrograph allows for the identification of deviations from the natural condition. The areas between the FDCs are called Ecodeficits (i.e. impaired FDC below natural FDC) and Ecosurpluses (vice versa). Summing all these areas for all 12 months of the year returns an overall alteration value in [m$^3$/s], which is to be minimized:

$$J_{\min}^{alt} = \sum_{month=1}^{12} ecodeficit_{month} + ecosurplus_{month} \quad (20)$$

**Experimental setup**

The general optimization framework is presented in Fig. 8. The optimization has been carried out by means of the freely available multi-objective optimization tool GODLIKE (Global Optimum Determination by Linking and Interchanging of Kindred Evaluators; Oldenhuis, 2010). GODLIKE uses the Non-dominated Sorting Genetic Algorithm II (NSGA-II; Deb et al., 2002) as the base. In each iteration, it randomly splits the population and allocates the subsets to the individual optimizers Genetic Algorithm (GA), Differential Evolution (DE), and Particle Swarm Optimization (PSO). Randomly interchanging population members between the algorithms increases the robustness of the optimization and the probability that the approximated Pareto set is close to the real efficiency frontier of the problem (Oldenhuis, 2010). It also reduces the need to fine-tune the parameters of the individual optimizers and makes GODLIKE applicable to a wide range of optimization problems. The idea of switching between individual optimizers was also adopted by other randomized search algorithms (e.g. "AMALGAM" by Vrugt & Robinson, 2007; "Borg" by Hadka & Reed, 2013). All these algorithms belong to a family of randomized search algorithms.

At every iteration of the randomized search, a generation of policies (i.e. each policy is a vector of 12 decision variable values) is generated and fed into the model. The model calculates the outcomes (i.e. time series $P_{total}$ and $Q_{total}$) of each of the policies by applying it to the model and running it over the modeling horizon. After that, the objective function values are calculated for each policy. Iterations continue until the stopping condition is met, i.e. when all policies in a generation are non-dominated and jointly form the Pareto approximate front.

The population size has been chosen to be quite high (i.e. 2500) for two reasons: firstly, to ensure a high resolution of the four-dimensional Pareto approximate front; secondly, to adequately explore the 12-dimensional decision variable space and mitigate the risk of premature convergence to local optima. The latter is especially important since GODLIKE does not include an automatic restart strategy during the optimization process. Extending GODLIKE with such a restart strategy could potentially improve the quality of the Pareto approximate set, however, this task was beyond the scope of this work. It is worth mentioning though that the mutation operator in GA and the potentially very large changes in velocities in PSO vectors occasionally result in large changes of individual population members, which helps to explore (previously unexplored) parts of the decision variable space.

To demonstrate the usefulness of the proposed method, the optimized policies must be validated on an "unseen" testing period. Therefore, the available data has been split into a training (2002-2011) and a test set (2012-2015); each of the two encompasses both very wet and very dry periods (see Appendix S2). The optimization on the training period has been carried out on a PC with 64 GB RAM and eight cores running at 3.30GHz. GODLIKE reported convergence after 175 hours, returning the Pareto approximate set of 2500 non-dominated policies. GODLIKE went through 123 iterations, running the operational model more than 300,000 times.

After the optimization, the resulting 2500 decision variables value vectors have been used as inputs for calculating the policies' performances over the testing period. As expected, the performances of the policies on the test set deteriorated if compared to the training set (e.g. the best performances decreased by 9% and 19% for firm and average energy, respectively). Naturally, some policies performed better than others on the "unseen" testing period (i.e. some are dominated by others), so that the resulting set was not Pareto-optimal anymore. An additional step of non-dominated-sorting identified a Pareto approximate set of 180 tested non-dominated policies.

## Results and discussion

The 180 policies of the tested four-dimensional Pareto approximate front show a great variety of operational strategies (Fig. 9). From these 180 policies, a decision maker can select a desirable trade-off, depending on his/her preferences and possible additional constraints. The observed system performance over the testing period (2012-2015) is indicated by the star in Fig. 9; exact performance values are listed in Table 4. Missing spill data in 2012 was simulated by fitting the model to the spill behavior of Porce III in 2013-2015.

When comparing the found optimal policies to the observed policy, the following has to be taken into account: the operators, in reality, do not optimize average and firm energy, but the revenue from energy generation. These objectives correlate but are not the same. This fact could not be taken into account directly when the OFs were formulated (mainly due to a lack of information regarding the business strategy of the operating company) and can be seen as a limitation of this study that will be further discussed in the conclusions. Therefore, a fair comparison between the observed and the optimized policies is unfortunately not fully possible, and the subsequent discussion is based only on the four objective functions defined in this study. However, the observed performance is still a useful reference for putting the quality of the optimized policies into perspective.

In Fig. 9, the arrows next to the labels of the four objectives illustrate their preference direction, which helps to identify the tradeoffs. The small range in the average energy performance (1062 MW – 1091 MW) indicates that this objective is only weakly in conflict with the other three. This can be attributed to a specific feature of this system: the powerhouses (except the smallest power plant Troneras) are not included in the dam structures, but situated further downstream and fed by the dams via diversion pipes, which creates high hydraulic heads. Therefore, the water level changes in the dams have only a limited influence on the overall hydraulic heads and the resulting amount of generated energy. Considering this, the increment in average energy compared to the observed performance (see Fig. 9 and Table 4) can mainly be attributed to the reduction of overall spillage, which is partly a result of the assumption of a perfect inflow forecast of one day. Fig. 9 and Table 4 also show that all 180 optimized policies perform better than the observed operation regarding flow regime alteration. This finding is not very surprising since the observed operation presumably takes flow alteration merely into account as constant MEFs.

Regarding the firm energy and flood hazard objectives, many optimized policies dominate the observed operation, too. The increment in firm energy shows the potential of adjusting operating rules to be better prepared for scarcity conditions, which are the periods when profit margins on the energy market are the highest. The reduction of the flood hazard demonstrates that it can be mitigated by explicitly including it into the operating objectives.

Table 4. Observed system performance and performance of the selected policy S over the testing period (2012–2015).

| Objective | Observed performance | Policy S |
|---|---|---|
| Firm energy (MW) | 729 | 796 |
| Average energy (MW) | 1,017 | 1,086 |
| Flood hazard (m$^3$/s·y) | 280 | 232 |
| Flow alteration (m$^3$/s) | 27.6 | 11.4 |

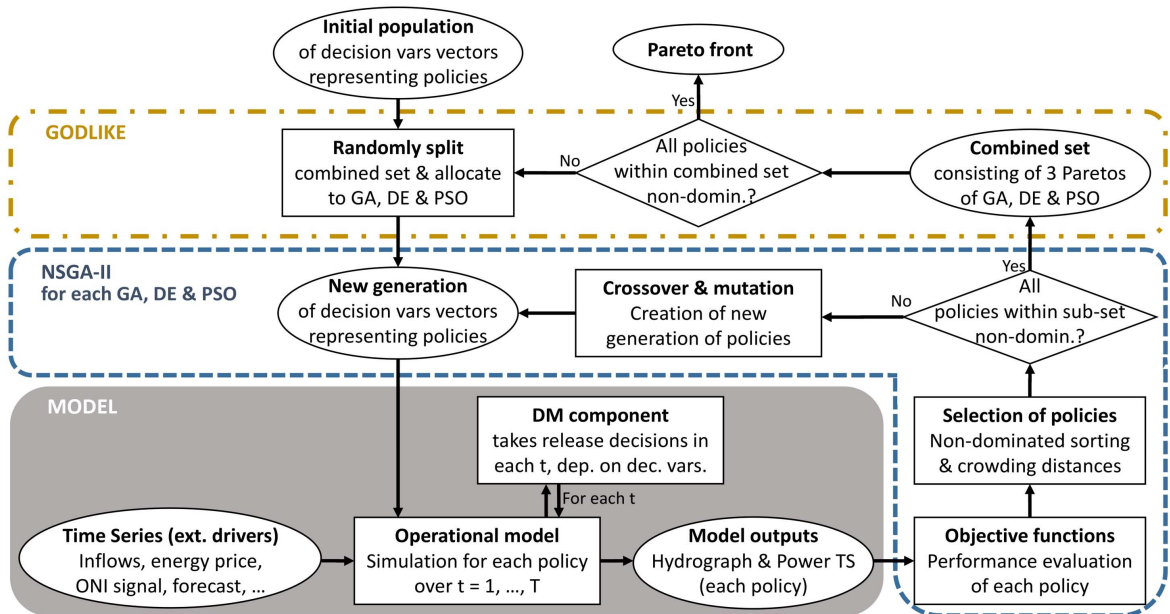

*Figure 8. Workflow of the multiobjective evolutionary algorithm and its connections to the model and decision-making component of the system.*

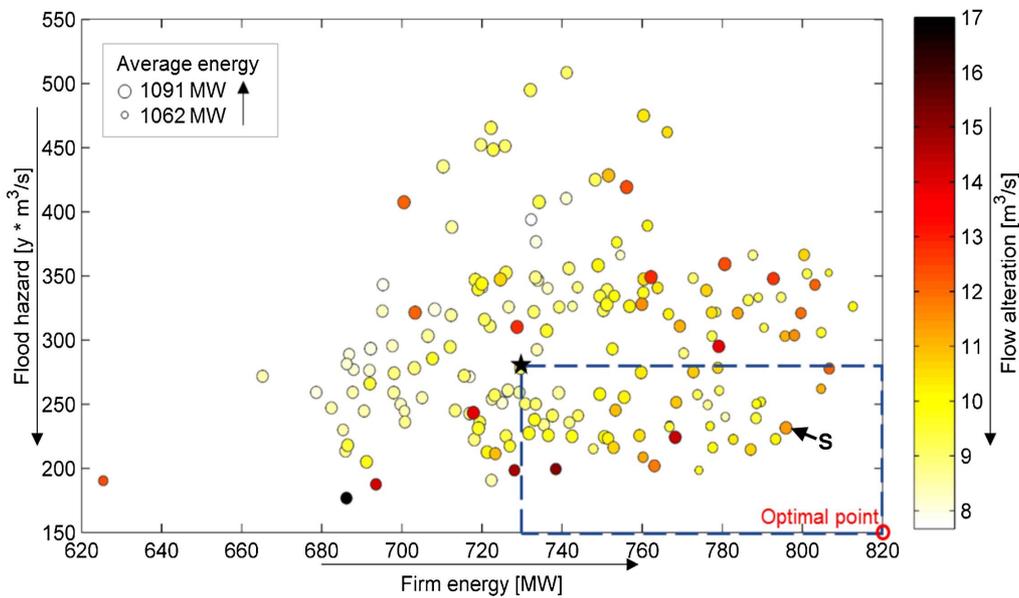

*Figure 9. Pareto-approximate set of operational strategies in the four-dimensional objective space. Flow regime alteration is shown in marker color, average energy generation in marker size. Arrows indicate the preference directions of the objectives. The observed performance over 2012–2015 is shown as the star. Policies in the dashed box dominate the observed performance, since they perform better with respect to all four objective functions.*

*Table 5. Summary of all 12 decision variables and their optimized values of policy S.*

| Purpose | Decision variables | Short description | Range | Policy S |
|---|---|---|---|---|
| G | $p$ | Weight decline toward future months (in calculation of $inflows_t$) | [0; 15] | 0.438 |
| | $v_1$ | Weight of weighted filling percentage ($filling_t$, system state) | [0; 1] | 0.637 |
| | $v_2$ | Weight of energy market price ($price_t$) | [0; 1] | 0.033 |
| | $v_3$ | Weight of persistence in monthly inflows ($inflows_t$) | [0; 1] | 0.168 |
| | $v_4$ | Weight of El Niño/La Niña influence ($enso_t$) | [0; 1] | 0.161 |
| A | $y_{drywet}$ | Inflow threshold between dry and wet season | [0; 250] | 134.7 |
| | $m_{P3}$ | Weight of low flow alteration rule (Porce III) | [0; 2] | 0.618 |
| | $b_{P2}$ | Target filling percentage of Porce II in wet season | [0; 1] | 0.603 |
| | $b_{P3}$ | Target filling percentage of Porce III in wet season | [0; 1] | 0.860 |
| | $a_{alloc}$ | Weight in allocation rule between filling-% and location of dams | [0; 1] | 0.375 |
| M | $r_M$ | Parameter in release function of Miraflores | [0; 23.6] | 1.733 |
| | $f_T$ | Target filling percentage of Troneras in dry season | [0; 1] | 0.915 |

Note: The first column indicates the purpose of each decision variable. G stands for calculation of the global generation target, A for target allocation rules and M for the release of Miraflores toward Troneras. $y_{drywet}$ has a double function and is also used for the release of Miraflores.

With respect to the four objectives defined in this study, the observed operation is dominated by a total of 45 optimized policies. Policy S (indicated in Fig. 9) has been selected for further analysis to demonstrate how operators can directly interpret the optimized physically-based operating rules. The values of the decision variables for policy $S$ are presented in Table 5. As a first step of the analysis, we focus on the weight distribution of the four components in the formula for the global energy target $GT$ (Eq. 2).

As expected, the weight $v_1 = 0.637$ of the system-state-component $filling_t$ is the highest of the four weights. The energy market price plays only a negligible role ($v_2 = 0.033$), which is hardly surprising. As mentioned earlier, in the optimization problem posed in this study the generated energy (in power units) was maximized, rather than the corresponding revenue, which is the main objective of the dam operators. In fact, the price component was only included in the study to set the stage for operators to adjust the objective functions to optimize revenue as opposed to generated power. This can be done by the operator, depending on the given energy market and the related confidential information at hand.

The remaining weights in the formula for the global energy target $GT$ are almost equal for the persistence in monthly inflows ($v_3 = 0.168$) and the El Niño/La Niña influence ($v_4 = 0.161$), which anticipate expected inflows in the future. The identified weight decline in the persistence in monthly inflows $p = 0.438$ (Eq. 5) results in weights [0.32, 0.28, 0.23, 0.17] for the current month and the three future months, respectively, giving future inflows relatively high importance. This moderate weight decline could also mean that the chosen approach of using historical inflow time series for estimating future inflows over the course of the year does not provide sufficient forecast skill, especially when taking into account that the historical inflow time series is also affected by El Niño/La Niña influence (see Fig. 2). However, the overall weight of the forecast component is not negligible ($v_3 = 0.168$), which reflects that it does contain useful information for decision-making, at least for policy S.

The allocation of the global generation target under market influence $GTM$ to the four individual dams with turbines is controlled by five decision variables (see Table 5). The threshold $y_{drywet}$ triggers dry and wet season rules. Its value of $y_{drywet} = 135$ m$^3$/s represents a trade-off between firm energy and flood hazard, retaining water and creating buffer capacities in advance of drought and flood periods, respectively. To give an idea of the meaning of this value: mean inflow is 173 m$^3$/s and 135 m$^3$/s corresponds to the 30$^{th}$ percentile of the flow-duration-curve. This means that flood buffer rules are triggered most of the time. The decision variable $y_{drywet}$ also has a second function in the control of the upstream drought buffer reservoir Miraflores (see DM Component Part 2). Its relatively low value limits the activation of the dry season condition to the driest periods and thus enables to target periods of water scarcity very precisely by the additional releases towards Troneras.

The flow alteration reduction rule is useful for reaching the outcome of Policy S, which is demonstrated by its weight of $m_{P3} = 0.637$. This value means that the minimum discharge requirement downstream of Porce III is in every month of the year fixed to 64% of the minimum ever observed discharge in the natural hydrograph in the corresponding month. Without a doubt, a higher value of $m_{P3}$ would further lower the flow alteration of policy $S$, however, this would be at the expense of the other objectives.

The target buffer capacities $b_{P2}$ and $b_{P3}$ in wet season bring a surprise. Porce III is controlled to lower its storage to 86% of its capacity (i.e. 17.8 million m$^3$ buffer) when flood buffer rules are triggered, indicating that it is not very important for flood protection, although it is the most downstream reservoir of the system. Porce II targets at lowering its storage to only 60% (i.e. 38.4 million m$^3$ buffer), suggesting that flood hazard originating from the southern branch is tackled best before the confluence with the northern branch. Again, it should be mentioned that even lower target levels in wet season would further reduce flood hazard, however, this improvement would not justify losses regarding the other objectives.

The allocation of the remaining target generation $GTM_{remaining}$ to the individual dams is controlled by $a_{alloc}$. Its value of $a_{alloc} = 0.375$ is surprisingly low since it suggests that the allocation of $GTM_{remaining}$ to the individual dams should be dominated by the relative position of the dams in the network, rather than by the individual filling percentages (see Eq. 13). This demonstrates that it can be a very useful measure to assign downstream dams with generally higher targets to lower their levels for flood protection.

Miraflores' role is providing the system with water in scarcity conditions. As described in DM Component Part 2, its release is controlled by the decision variables $r_M$, $y_{drywet}$ and $f_T$. The comparably low value of $r_M = 1.73$ m$^3$/s (see Fig. 5) confirms that in the wet season, almost all water is retained in the dam and only released when Miraflores runs into danger of being completely filled. In the dry season, the target filling $f_T$ in the receiving dam Troneras, dominates the release decision. Its comparably high value of 92% ($f_T = 0.915$) indicates that in the dry season, the diversion channel between the two reservoirs is almost continuously at its maximum capacity. This confirms the initial impression that the dry season is targeted very precisely, as the low value of $y_{drywet}$ classifies only the driest periods as dry season. Providing the required amount of water to maintain the filling of 92% in Troneras would not be possible over longer periods of the year, as Miraflores would be completely emptied at some point.

## Conclusions

This paper presents a multi-objective optimization approach of DPS that addresses the gap between existing multi-objective reservoir optimization methods and the physically-oriented perspective of dam operators, with the aim to overcome policy myopia. The specific feature of this study is that this is done by employing parameterized physically-based operating rules, instead of the parameterized black-box functions (data-driven models, e.g. RBF), which were used in the earlier studies related to DPS.

The concept has been applied to a network of five hydropower dams in the Nechí catchment in Colombia, involving four objectives: average and firm energy generation, and the reduction of flood hazard and hydrologic flow alteration. A detailed operational model of the system has been built; the daily release decisions are taken by applying six parameterized physically-based operating rules. These rules have been developed depending on the physical characteristics of the given case study and are thus not directly transferrable to other systems, but the general approach of developing and optimizing such rules is. Furthermore, some of the rules (e.g. all of the target allocation rules and some components of the global generation target rule; see Table 2) can serve as a base for developing similar rules for other reservoir networks and require only adjustments to the given physical characteristics and data.

The six physically-based operating rules depend on only 12 decision variables. Controlling a network of five storages using such a low number of decision variables significantly reduces computational complexity compared to prior DPS approaches. For instance, Biglarbeigi et al. (2014) used 117 parameters to optimize the release policy of a network of three reservoirs. Giuliani et al. (2018) used 176, and Quinn at al. (2018) 134 parameters for a system consisting of four reservoirs. Unfortunately, computational limitations impede the possibility to explore quantitatively how much the proposed method of introducing physical interpretability sacrifices in performance, compared to conventional DPS. However, given the adequate computational resources are available, such a comparison would be important to carry out. The limited computational resources also impeded a more complex experimental setup of the optimization. For instance, instead of using only non-dominance of all population members as the stopping criterion, the convergence towards the Pareto-approximate front could be monitored by employing a measure of search progress, such as hypervolume. Besides, multiple seeds could be run to ensure a consistent algorithm performance.

Four objective functions have been considered. The system model and its decision-making component have been coupled with a multi-objective evolutionary algorithm. The simulation-based optimization approximated a Pareto front of general operating policies. The performance of the optimizes policies has been compared to the observed performance of the system over a testing period of four years. With respect to the four objectives defined in this study, many of the optimized policies dominate the observed performance over the testing period; especially the improvements regarding flood hazard and flow regime alteration are considerable. It is important to point out that the observed performance is a result of the operator's objective to maximize revenue from energy generation, while in the current study mere power units are optimized (these objectives correlate but are not the same). Therefore, a fair comparison of optimized and observed energy performances was not fully possible. Operators, who have more information on the energy market, could address this by adjusting the objective functions to evaluate revenue instead of power units. In the scope of this work, it was unfortunately not possible to carry out this complex task, mainly due to insufficient information about the operator's strategy on the energy market. It is planned to address this limitation in future work.

Although the testing period contained both very wet and very dry periods, it would be important to test the optimized policies under an even higher hydroclimatic variability, for instance by using synthetic inflow time series. This was unfortunately out of the scope of this paper, as the system is fed by seven separate inflow times series. However, this might be addressed in the further development of the approach, and the optimized policies will be in any case also further tested on more recent inflow data.

The analysis of one of the optimized policies demonstrates how the results could be directly used and interpreted by system operators. It shows that the presented framework offers operators an approach to overcome policy myopia from their perspective and adjust their operating rules to perform better, facing the evolving demands and interests. Future work will aim at further improving the reflection of the decision-making practice in the formulation of the operating rules and the objective functions. To this point, the full-fledged real-life evaluation of the presented method by the system operators was unfortunately not possible. However, negotiations for a collaboration with the operating company EPM are on-going and will hopefully lead to further improvement of the framework and to its adaption in the working practice.

## Data Availability

The input time series used in this study are freely available online (XM Compañía de Expertos en Mercados, 2017). For technical data on the reservoirs and turbines, please contact the operating company Empresas Públicas de Medellín (EPM). The MATLAB computer codes and data generated in this study are freely available on GitHub (Ritter, 2017).

## Acknowledgments

The first author acknowledges the EC Erasmus Mundus Program (MSc in Flood Risk Management). The Nature Conservancy supported the first author in his investigations in Colombia. Our gratitude is extended to the Colombian Institute of Hydrology (IDEAM), the Escuela Colombiana de Ingeniería, and the site engineers of EPM for the provided data and fruitful discussions.

## Supplemental Material

The Supplemental Material is available online in the ASCE Library (https://ascelibrary.org). It includes a) details about the extension of the inflow time series of the dam Porce III, and b) the system inflow hydrographs for the training and validation periods.